# Frequency Independent Framework for Synthesis of Programmable Non-reciprocal Networks

Ruochen Lu[1], John Krol[1], Liuqing Gao[1], and Songbin Gong[1]

[1]University of Illinois at Urbana Champaign

**Abstract:**

Passive and linear nonreciprocal networks at microwave frequencies hold great promises in enabling new front-end architectures for wireless communication systems. Their nonreciprocity has been achieved by disrupting the time-reversal symmetry using various forms of biasing schemes, but only over a limited frequency range. Here we demonstrate a framework for synthesizing theoretically frequency-independent multi-port nonreciprocal networks. The framework is highly expandable, and can have an arbitrary number of ports while simultaneously sustaining balanced performance and providing unprecedented programmability of non-reciprocity. A 4-port circulator based on such a framework is implemented and tested to produce broadband nonreciprocal performance from 10 MHz to 900 MHz with a temporal switching effort at 23.8 MHz. With the combination of broad bandwidth, low temporal effort, and high programmability, the framework could inspire new ways of implementing multiple input multiple output (MIMO) communication systems for 5G.

**Introduction:**

Microwave frequency nonreciprocal networks that bear non-reciprocal response have long been sought after for a wide range of applications, including full-duplexing radios[1,2] and quantum computing[3,4,5]. Most commonly utilized nonreciprocal multiport networks are isolators and circulators. Conventionally, non-reciprocity is obtained by magnetically biasing a ferrite material with in which the electromagnetic wave propagates at different phase velocities in the opposite directions[6,7]. In a circular structure based on a material of such properties, constructive and destructive interference of the clockwise and counter-clock wise propagating waves can exist at different nodes around the circular resonator, thus establishing transmission and isolation through ports situated at these nodes.

Motivated by attaining non-reciprocity for more integrated RF and microwave applications, temporal modulations, applied to either reactive[8,9,10] or conductive[11,12] elements, have recently been explored to produce a momentum-biasing equivalent to the magnetic ones and break the reciprocity. These approaches all rely on wave interference or mode splitting caused by biasing in a resonant structure. In other words, the bandwidth over which their desirable non-reciprocal performance can be maintained are sensitive to phase delays between adjacent ports of the network. Although wide-band phase nonreciprocal gyrators[12] can be engineered to enhance the bandwidth of such networks, these type of non-reciprocal devices are inherently frequency dependent. Moreover, demonstrations on temporally modulated nonreciprocal so far are primarily two port gyrators[11,13] and three port circulators[14]. Conceivably, both magnetic and temporal modulation based approaches can be expanded to a network with more ports by exploiting established circuit topologies or simply networking several 3-port circulators. However, the possibilities of reconfiguring the non-reciprocity in these approaches are limited. For instance, only a small subset of circulation sequences through all ports are accessible among all permutations, due to the limitations arising from its topology and application of momentum biasing.

We show a framework for synthesizing a frequency independent and broadly programmable non-reciprocal network with an arbitrary number of ports (2N) using switches and an array of dispersionless delay lines. The generalized 2N-port framework can also be elegantly reduced to 3 port or 2 port device with more compact size and less switched delay lines than the sequentially switched delay lines[15]. This concept attains multi-port non-reciprocity by equally multiplexing the input signal onto N delay lines in the time domain and later aggregating the delayed signals off N delay lines consecutively at the intended port. The timing offset between switches addressing each port results in only one port receiving signal at any given time from an excitation port. Unlike the abovementioned momentum biasing approaches, the non-reciprocal performance of our network is only dependent of the time delays, instead of phase delays, and therefore is frequency independent. More impressively, the network has far more programmable states than any alternative reconfigurable non-reciprocity. Such programmability of nonreciprocity in a multi-port network will inspire new applications in multiple input multiple output (MIMO) communication systems.

**Results:**

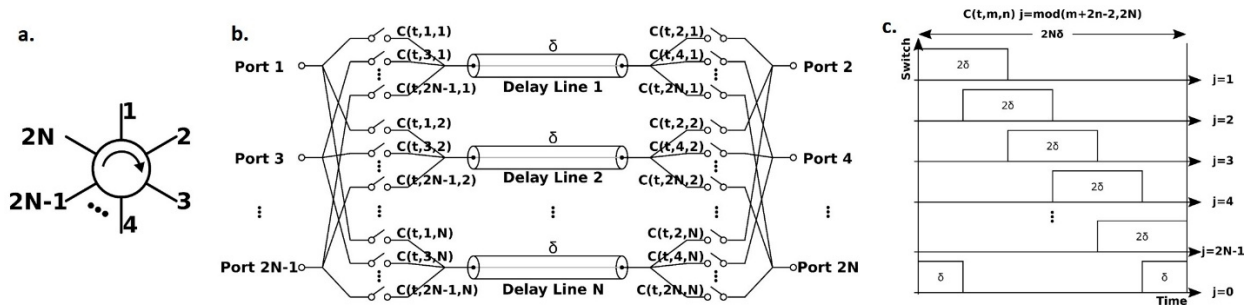

Fig. 1. a. Schematic symbol of a circulator of 2N ports and clockwise circulator. b. Concept of 2N-port non-reciprocal network. c. Switch control waveforms applied to the network for producing the nonreciprocity. d. 3-port circulator derived from the 2N framework

Fig. 1 shows the 2N-port framework consisting of 2N ports equally situated on the both sides of $N$ identical delay lines that each has a time delay of $\delta$. On either side of the delay lines, each port is fanned out to connections with all delay lines through a single pole single throw (SPST) switch that presents open in the off state. Therefore, composition of such a 2N-port network require N delay lines and $2N^2$ SPSTs (or 2N SPNTs).

The clock signal for controlling each switch is denoted as $C(t,m,n)$, where $t$ is the time, m is the port number, n is the delay line number. All the clocks have a period of $2N\delta$, and a duty cycle of $1/N$. Within the time range [0, $2N\delta$], the control signal can be represented as:

$$C(t,m,n) = \begin{cases} H[t-(j-1)\delta] - H[t-(j+1)\delta] & \text{for } j \neq 0 \\ H[t] - H[t-\delta] + H[t-(2N-1)\delta] - H[t-2N\delta] & \text{for } j = 0 \end{cases}$$

where $H$ is the Heaviside step function, and j is the remainder of the modulo operation.

$$j = \mod(m + 2n - 2, 2N)$$

*C(t,m,n)* is designed to turn on only one switch, among the switches connected to Port m, at any given time so that the signal is sequentially time-multiplexed onto the N delay lines. On the other side of delay lines, Port m+1 is controlled by C(t,m+1,n), which is designed to be a time delayed version of C(t,m,n) with a timing offset of δ so that the signal, after traversing N delay lines, will be collected and de-multiplexed into Port m+1.

In the reverse path, signals fed into Port m+1 , after being time multiplexed onto and traversing the delay lines, are subsequently rejected by port m because the switching control clocks, C(t,m,n), are the time advanced version of  C(t,m+1,n).  In other words, all switches are turned off as signal arrives Port m from Port m+1. On the other hand, switches on Port m+2 are synchronized with the arrival of signals from Part m+1 to aggregate them from the delay lines. The exception exists for Port 2N, to which the fed signals will be circulated to Port 1.

For a 2N-port network that consists of infinitely fast and lossless switches and lossless dispersionless delay lines, and is addressed by ideal square wave control signals, infinitely large isolation, zero insertion loss, and zero return loss in the circulation are predicted. The perfectly synchronized time-domain multiplexing and de-multiplexing on opposite ends of the N delay lines allow signal incident from Port m to exclusively transmit to Port m+1, while the energy leakage in the reverse order is completely forbidden.

Note that in our generalized framework, N has to be an even number as required by the symmetry of the network. For producing odd number of ports, a network with an even number of ports can be reduced to have one less port by leaving one port open, which essentially eliminates N SPSTs. As an example seen in Fig. 2a and 2b, a 4-port network is reduced to a 3-port circulator that is typically sought after for full-duplex radio applications.

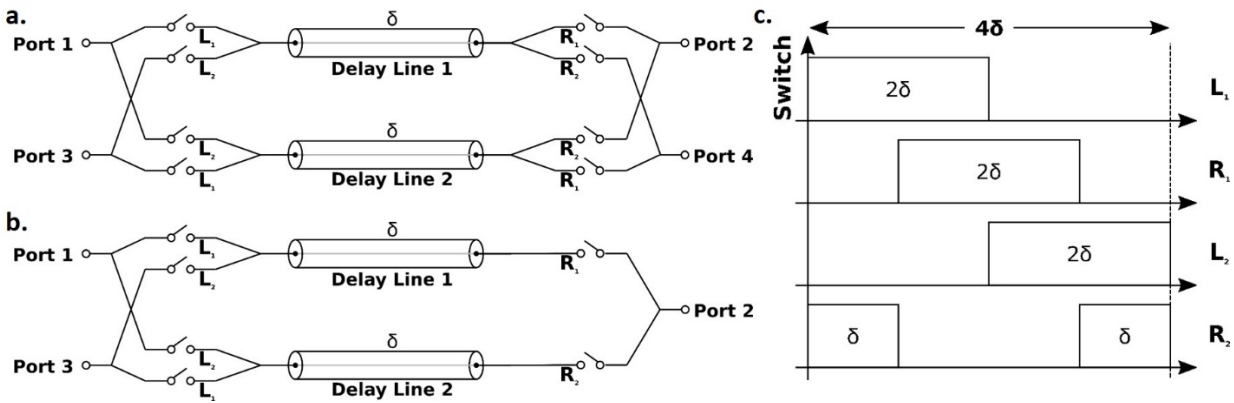

Fig. 2. a. Schematic of a 4-port circulator. b. schematic of a 3-port circulator reduced from a 4-port circulator. c. Switch control waveforms for producing clockwise (from 1 to 4) circulation.

**4-port Broadband Circulator and Experimental Validation**

To experimentally validate our framework, we choose to produce a 4-port circulator based on the 2N port framework with a frequency span from DC to 1 GHz. Fig. 2 shows the schematic and control waveforms.

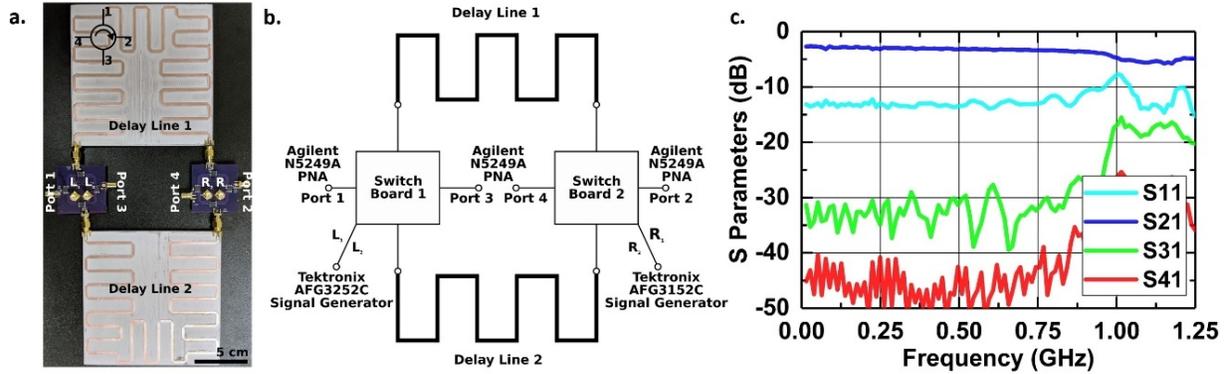

Fig. 3. a. Picture of the implemented 4-port circulator consisting two switching modules and two microstrip delay lines. b. Block diagram of the constructed system. c. Simulated S-parameters of the 4-port circulator.

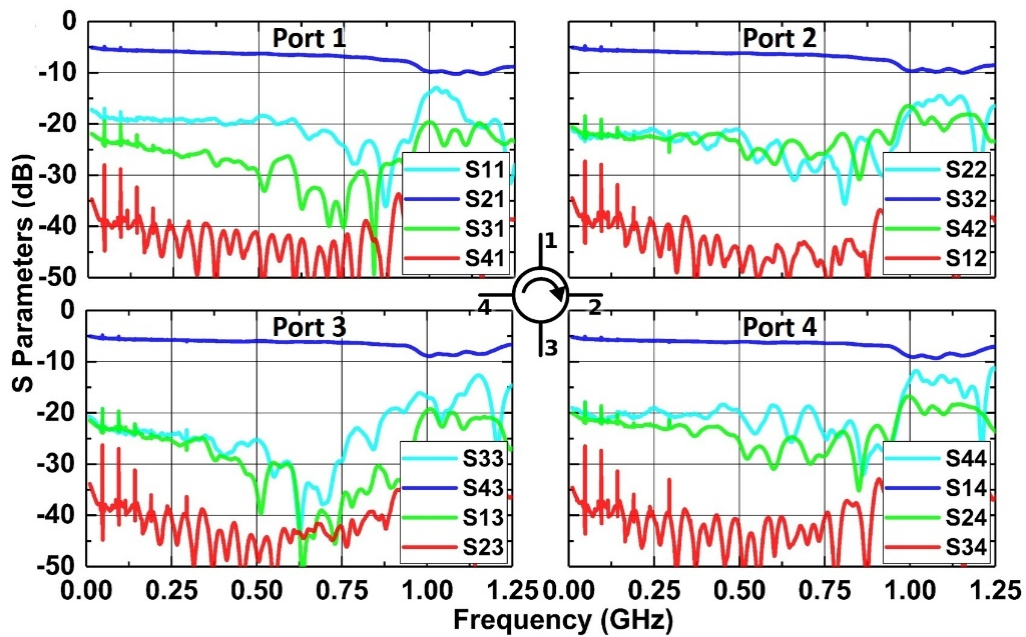

Fig. 4. Measured S-parameter performance of the 4-port circulator.

As seen in Fig. 3(a), the prototype is implemented with connectorized switching and delay line modules. Two delay line modules, with each end connected to a switching module, form the nonreciprocal network. We take the modular approach for experimental validation as it allows more nodes to experimentally observe performance and analyze loss in the 4-port network.

Based on our design for a 4-port switching module, only 2 single pole single throw series switches that present near open circuit to the input in the off-state are needed. In practice, open-reflective switches are not commonly available with fast switching time. Alternatively, 4 short-reflective switches, Minicircuit MSW 2-20+, are arranged in a lattice configuration (see supplementary materials) to equivalently produce switching of two SPST open-reflective switches. MSW 2-20+ has a fast switching time of 2 nS, which minimizes the insertion loss due to switching. The delay line modules are implemented using Roger Duroid 6010.2LM boards with meandering microstrip structures to produce a total group delay of 10.5 ns with slight dispersion that is less than 1 ns.

In operation, the switches are controlled by 4 clock signals that have a period of 42 nS, and a frequency of 23.8 MHz. The slightly increased delay is caused by the additional electrical length in the control boards. The switches on the same side of the delay line are complimentarily driven while the switches on the opposite ends of the same delay are driven with a timing offset of 10.5 nS. The clock signals are generated by two synchronized dual-channel Tektronix arbitrary function generators, and fed to the control ports on the switching modules.

Advanced Design System (ADS) is used for simulating the 4-port performance. The switches have 2 ns switching time, an on-state resistance of 3Ω, and an off-state resistance of 60 kΩ. The delay lines are represented by their S-parameter performance, which is modeled using ADS momentum. To extract the frequency domain response of the network, a series of time domain simulations with varying single tone inputs to Port 1 are performed before Fourier transform is performed to attain scattered power out of other ports at the input frequency. As seen in Fig. 3c, the simulation shows a broadband (up to 0.9 GHz) nonreciprocal performance. An IL of 3 dB at low frequencies is caused by the non-ideal switch properties and the loss in the delay lines. A great isolation over 30 dB is observed simultaneously. The performance degrades at higher frequencies due to the additional loss in the delay lines.

The measurement of 4-port was done using a setup shown in Fig. 3b. The non-reciprocal network is tested with a 4-port Keysight PNA-X network analyzer. Calibration is performed with Keysignt 85052D calibration kit to move the measurement reference planes to the connectors on the switch modules. 4-port S-parameter is subsequently characterized with IF bandwidth of 1 kHz and a measurement power level of −5 dBm.

As seen in Fig. 4, broadband non-reciprocal responses are obtained from 10 MHz to 0.9 GHz. A minimum IL of 5.1 dB is obtained at low frequencies. Isolations of 35 dB is measured between the adjacent ports, and 20 dB between the diagonal ports. As the frequency increases to the self-resonance in the delay lines around 0.9 GHz, the IL and isolation performance gradually decays to 7.6 dB and 24 dB, respectively. The measured performance slightly deviates from the simulated results due to the simplification of switches in the model and multi-reflections on delay lines caused by impedance mismatch at ports.

**Discussion**

- Frequency Independent Performance

As discussed earlier, the frequency independent performance of nonreciproicity is the outcome of perfect synchronization of time-domain multiplexing and delays in the forward path, and complete off-synchronization between them in the backward route. We recognize that a number of causes in practice can compromise the frequency independent performance and yield a broadband performance instead. For instance, the electromagnetic delay lines typically exhibit dispersion, which causes the synchronization between switching and delay to degrade as the operating frequency moves off the design center frequency. To reduce size, delay lines based on slow-wave or meandering structures often have a cut off frequency that also limits the BW of the nonreciprocal network. Other types of delay lines with smaller sizes, e.g. acoustic delay lines[16], usually have passbands over which low insertion loss and constant group delay can be maintained. Nonetheless, with our frequency independent framework as the basis, the bandwidth over which non-reciprocity is enabled should be only limited by the components chosen for implementation.

It is worth noting that the frequency independent performance is not dependent on the temporal effort applied in the system. Unlike the momentum biasing approaches where the bandwidth of nonreciprocity is fundamentally limited by the modulation frequency used to produce momentum biasing, the switching frequency in our framework is only set by the time delay length imposed by the delay lines. Provided with low loss delay lines to render long group delays, the switching frequency can be reduced to a mere fraction of the non-reciprocal bandwidth (e.g. 23.8 MHz switching frequency for maintaining a nonreciprocal bandwidth of 900 MHz in our case), consequently giving rise to simpler and lower cost clock generation, less phase delay in clock signal fanout, and minimized overall temporal effort. One caveat in operating our framework lies in the resulting group delays between ports, which is longer than those of ferrite circulators. Therefore, such systems might not be a good fit for timing-sensitive applications (e.g. radar front ends).

- Network expandability without compromising performance and symmetry

Expanding a momentum-biased 3-port circulator into an N-port circulator is a non-trivial task. Simply adding more folds of symmetry in the structure will not produce unilateral circulation. In other words, the excitation at one port will be nonreciprocally received at more than one port. A typical way to attain nonreciprocal networks with more ports using momentum-biased devices is to network 3 port circulators in various manners, such as the method reported for creating macroscale topological materials[17]. With each added circulator in the network, the number of ports in the network can be maximally increased by one, thus suggesting a heavy cost in component counts and clock feeds for constructing multi-port nonreciprocal networks beyond 3 ports. Additionally, networking 3-port circulators often breaks the network structural symmetry and creates unbalanced paths between ports. Consequently, higher insertion loss are expected for paths that require the signal to traverse more in the composed multi-port network to reach destination ports.

For the even-port operation in our time-multiplexed framework, one can add 2 more ports to the network with each added delay line, which compares favorably against the network expansion via interconnecting 3 port circulators. For networks with an odd number of ports, the cost of expansion is the same, except for adding the last port, which requires a delay line for its own. More importantly and more advantageously in our framework, all transmission paths are balanced with the same insertion loss and delay regardless the number of ports. Thus, the 2N-network maintains N folds of symmetry in both the structural design and performance.

- Programmability of nonreciprocity with a rich space of permutations

Enabling programmable RF circuits has been the holy-grail problem for designing highly adaptive RF systems in the past decade, focusing primarily on either passive reciprocal networks, such as filters[18,19,20,21], antenna tuners[22], and phase shifters[23], or active/nonreciprocal circuits, such as amplifiers[24].

Programmability of passive non-reciprocity has rarely been visited even though the current carrier aggregated communication systems can greatly benefit from programmable non-reciprocity in front ends[25]. Temporal modulated non-reciprocal systems have recently revived the hope for achieving such programmability without compromising other relevant performance specifications.

Our framework is readily programmable by first re-shuffling the clock waveforms applied to the switches on one side of the delay lines, and then adjusting the clocks on the other side accordingly. Through this practice, any port on one side of the delay lines can be configured to circulate to any port on the other side of the delay lines, thus allowing for a rich space of non-reciprocal states. The accessible states for the 2N-port nonreciprocal network can be studied as S-matrix permutations with the only limitation that circulation between ports on the same side of the delay lines cannot be established. Therefore, assuming all ports are matched, the components in the shaded regions of the S-matrix, seen in Fig. 5a, are inaccessible for programming. On the other hand, assuming the network is lossless and S-matrix is unitary, the sub-matrices outlined by the red boxes in Fig. 5a have a single complex component in each row and column. Provided that the implementation is balanced with identical switches on both sides of the identical delay lines, these complex components are identical with a magnitude of 1, and are denoted as αs. Note that the programming of the network changes neither the structural nor the performance symmetry. In other words, the programming does not change the value of α in the S-matrix.

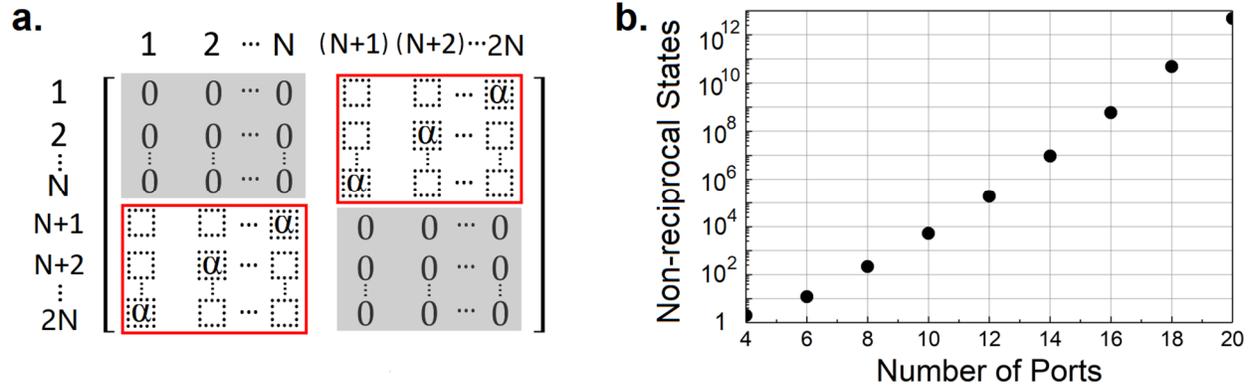

Fig. 5. a. Accessible and forbidden regions of the S-Matrix for programing. b. Programmable non-reciprocal states as a function of number of ports.

To determine the number of programmable non-reciprocal states, we can first populate the top right sub-matrix, referred as sub-matrix A onward, with allowed permutations, which is $N!$. With each permutation of A, we can then exam the allowed permutations of sub-matrix B in the lower left quarter. Due to non-reciprocity of the network ($S_{ij} \neq S_{ji}$), $N$ components are determined as 0 in B for a given permutation of A. Consequently, the number of ways to populate B for a given A is given by:

$$P(N \times N, N) = N! - C_N^1 \times (N-1)! + C_N^2 \times (N-2)! - C_N^3 \times (N-3)!$$
$$+ \cdots + (-1)^{N-1} \times C_N^{N-1} \times (1!) + (-1)^N \times C_N^N$$

Thus, the number of nonreciprocal states, Ω, for a 2N-port network is:

$$\Omega\,(2N) = (N!) \times \mathrm{P}(N \times N, N)$$

As seen in Fig. 5b, this represents an exponential growth of programmable non-reciprocal states as the number of ports increases.

**Method:**

The loss in the system can be understood with an analytical approach focusing the switching loss, which is defined as the insertion loss caused the switching process. Thus, when analyzing switch loss, the delay lines are modelled as lossless and perfectly matched transmission lines. Fundamentally, the switch loss is the result of momentarily losing signal during the switching from one delay line to the other. Such a loss is inevitable using switches with small but not zero switch on and off time. The Insertion loss due to switching is determined by how much signal is lost proportionally over time, and thus related to ratio of switch time ($t_s$) to delay time ($\delta$). The switches are represented as time-varying resistances ($R_{switch}$) during switching on and off periods. They linearly change resistance from an off-state resistance ($R_{off}$) to on-state resistance ($R_{on}$) over a switching period ($t_s$) upon the application of control waveforms, which are assumed to be perfect square waves of 50% duty cycle. In a $2\delta$ period, $R_{switch}$ can be described as:

$$R_{switch}(t) = \begin{cases} R_{off} + (R_{on} - R_{off}) \cdot \dfrac{t}{t_s} & \text{for } 0 < t \leq t_s \\ R_{on} & \text{for } t_s < t \leq 2\delta - t_s \\ R_{on} + (R_{off} - R_{on}) \cdot \dfrac{t - 2\delta + t_s}{t_s} & \text{for } 2\delta - t_s < t \leq 2\delta \end{cases}$$

Consider the upper line in Fig. 2a, the input signal from $0 < t < t_s$ experiences a time varying transmission coefficient of $h(t)$ when transmitting through the switch controlled by $L_1$. Then, this signal is delayed by $\delta$, and from $\delta < t < \delta + t_s$ the signal experiences a transmission of $h(t - \delta)$ when transmitting through the switch controlled by $R_1$. Given that switching time ($t_s$) is smaller than $\delta$, $h(t)$ can be described as:

$$h(t) = \frac{2Z_0}{R_{switch}(t) + 2Z_0} \qquad \text{for } 0 < t \leq 2\delta$$

The transfer function, between Port 1 to 2 as seen in Fig. 2a, is given as:

$$H_{sys}(\omega) = H(\omega) \cdot e^{i\delta\omega} \cdot H(\omega - \delta) = H(\omega)^2$$

Where $H(\omega)$ is the Fourier transform of $h(t)$. It is noteworthy that when $t_s>0$, the system transfer function has components other than the DC component. It implies that the non-ideal switching produces signals at frequencies other than the input signal (e.g. the carrier frequency), which is another interpretation of the switching loss. In addition, insertion loss is also introduced by by $R_{on}$ and $R_{off}$. Thus, the total insertion loss (*IL*) between ports can be described as:

$$IL = -20\log_{10}(v_{out}/v_{in}) = -20\log_{10} H_{sys}(0)$$

Based on the analytical closed form expression of switching loss as a function of $t_s$ and $\delta$, a 2D contour plot of switching loss with switching time varying from 0 to 5 ns, and group delay of delay lines varying from 10 to 50 ns is plotted in Fig. 6. An $R_{on}$ of 6 Ω and an $R_{off}$ of 120 kΩ are assumed for the switches used in implementation.

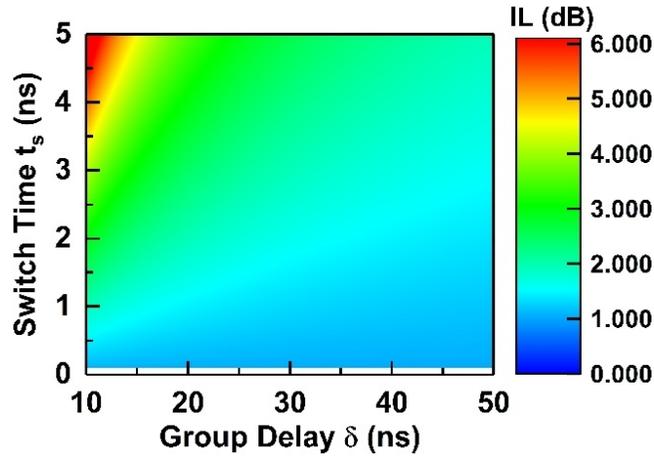

Fig. 6. Dependence of switching loss on switching time and group delay.

**Data availability**

All relevant data is available upon request.

**Acknowledgements**

This work is partially supported DARPA MTO signal processing at radio frequency program (SPAR) program under grant number HR0011-17-2-0004.

**Author contributions**

R. L., J. K. and S. G. conceived the ideas for the frequency independent nonreciprocal networks. R. L. J. K. and L. G. developed the theoretical model for predicting the performance. R. L. designed, implemented, and tested the switching modules. L. G. designed, implemented, and tested the microstrip delay lines. R. L. assembled the 4-port circulator, performed and analyzed the measurements. R.L., L.G. and S.G. wrote the manuscript. S. G. supervised the research.